\documentclass[aps, prl, twocolumn, superscriptaddress, showpacs]{revtex4}

\usepackage{graphicx}

\begin{document}
\title{Source coding by efficient selection of ground states clusters}

\author{Demian Battaglia}
\affiliation{SISSA, Via Beirut 9, I-34100 Trieste, Italy,}

\author{Alfredo Braunstein}
\affiliation{SISSA, Via Beirut 9, I-34100 Trieste, Italy,}
\affiliation{ICTP, Strada Costiera 11, I-34100 Trieste, Italy.}

\author{Jo\"el Chavas} 
\affiliation{ICTP, Strada Costiera 11, I-34100 Trieste, Italy.}

\author{Riccardo Zecchina} 
\affiliation{ICTP, Strada Costiera 11, I-34100 Trieste, Italy.}

\begin{abstract}
In this letter, we show how the Survey Propagation algorithm can be generalized
to include external forcing messages, and used to address selectively 
an exponential number of glassy ground states.
These capabilities can be used to explore efficiently the space of solutions
of random NP-complete constraint satisfaction problems, providing a direct 
experimental evidence 
of replica symmetry breaking in large-size instances.
Finally, a new lossy data compression protocol is introduced, exploiting as a 
computational resource
the clustered nature of the space of addressable states.
\end{abstract}

\maketitle

The combinatorial problem of satisfying a large set of constraints
that depend on N discrete variables is a fundamental one in
computer science (optimization and coding theory) as well as 
in statistical physics (frustrated systems), and may become
extraordinarily difficult even for randomly generated problems as soon as some
control parameters are selected in specific ranges.
While in general the study of the connection (if any) between worst-case and 
typical-case computational complexity of hard (NP-complete \cite{Garey_Johnson}) 
CSPs 
has not yet been fully  developed, recent advances in the statistical mechanics 
study of 
random constraint satisfaction problems (CSPs) have connected
the origin of such computational intractability to the onset of clustering in
the space of optimal assignments and to the associated 
proliferation of metastable states \cite{Nature, TCS}.
An important  byproduct of the analytical studies of random
CSPs has been the introduction of a new class of
algorithms ---the so called Survey Propagation (SP)  algorithms 
\cite{SP, BMZ}--- specially devised to deal with the clustering scenario and able 
to find optimal assignments of 
benchmark problems on which all known optimization 
algorithms fail.

In this letter we make a step forward in understanding the potentialities
of Survey Propagation techniques by answering two questions:

(i) Is it possible to explore efficiently the space of solutions
of a given random  combinatorial problem ? 

(ii) Can we use this capability of addressing a  large
 set of states for computational purposes?
 
The first question relates to the physical issue of probing exactly the topology 
of the 
space of solutions (ground states) in problems that are in a clustered phase (the 
so 
called replica symmetry breaking (RSB) phase). Surprisingly enough, such 
geometrical insight is
also important for engineering applications in information theory (\textit{e.g.} 
LDPC error correcting codes \cite{LDPC}).
The second question addresses a new algorithmic perspective in which the 
presence of many states becomes a powerful resource of the computational device.

In what follows we shall provide a positive answer to both questions
by providing an efficient generalization of SP which is indeed capable
of addressing efficiently -- computational cost almost linear in the
size of the problems -- an exponential number of different clusters of
ground states that are invisible to the known search algorithms.  From
the physical side, we provide the first numerical evidence for the RSB
geometric structure for large size instances of NP-complete problems.
On the computational side we show that one can indeed take advantage
of the addressability of the set of clusters of ground states to
produce a ``physical'' lossy data compression scheme \cite{mackay}
with non-trivial performance.

A generic constraint satisfaction problem is defined by $N$ discrete
variables ---\textit{e.g.} Boolean variables, finite sets of colors or
Ising spins--- which interact through constraints involving typically
a small number of variables. The energies $C_a(\vec{\xi})$ of the
single constraints (equal to 0 or 1, depending if they are satisfied
or not by a given assignment $\vec{\xi}$) sum up to give the global
energy function $\mathcal{E}$ of the problem and are function of just
a small subset of variables $V(a)=\{j_{a_1}, \ldots, j_{a_K}\}$ (every
variable $j$ is involved on the other hand in a subset $V(j)$ of
constraints).  Since one is interested in satisfying all the
constraints simultaneously, the CSP is just equivalent to the problem
of looking for zero energy ground states. For most NP-complete CSP the
function $\mathcal{E}$ can be directly interpreted as a
spin-glass-like hamiltonian \cite{TCS, SP, coloring}.  For instance,
the well known case of the random K-SAT problem consists in deciding
if $M$ clauses ---taking the form of the OR function of $K$ variables
chosen randomly among $N$ possible ones--- can be simultaneously true.
The energy contribution associated to a single clause can then be
written as $C_a(\vec{\xi})=\prod_{l=1}^K \frac{1}{2} (1+J_{a,
  l}x_{a_l})$, where $x_{a_l}=\pm 1$ depending on the truth value of
$j_{a_l}$ and $J_{a, l}=\pm1$ if $j_{a_l} $ appears negated or
directed in the clause $a$.  The same variable can appear directed and
negated in different terms and hence give rise potentially to
frustration.

\begin{figure}[t]
\begin{centering}
\includegraphics[scale=0.19]{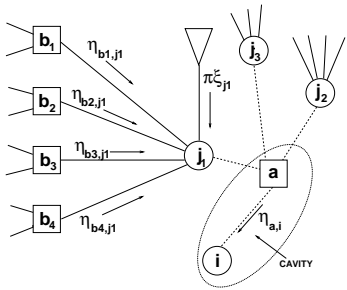}
\caption{A portion of factor graph}\label{1}
\end{centering}
\end{figure}

The connectivity distribution and the loop structure of the
\textit{factor graph} associated to a CSP (in Fig.~\ref{1}, variables
are represented as circles connected to the constraints in which they
are involved, depicted as squares) has a strong influence on the
behavior of search algorithms.  For many important random CSPs, when
the ratio $\alpha=\frac{M}{N}$ is included in a narrow region
$\alpha_d < \alpha < \alpha_c$ ---the exact values of the thresholds
depending on the details of the problem and on the chosen random graph
ensemble --- the problem is still satisfiable but the zero energy
phase of the associated hamiltonian breaks down in an exponential
number of clustered components.

The cavity method of statistical physics, used for accurate analytical
computations \cite{SP, MMZ, MoPR} of the threshold locations in good
agreement with the numerical experiment \cite{kirkpatrick:selman:94},
provides as well the theoretical foundation of the Survey Propagation
(SP) message passing algorithm, successful in the resolution of
instances of both the $q$-coloring and the $K$-SAT problem
\cite{coloring, SP} which are hard for local search algorithms.  A
constraint node $b$ is supposed to send a message $\vec{u}_{b\to
  j}=\vec{e}_s$ (where $\vec{e}_s$ are vectors having just the $s$-th
component equal to 1) to a variable $j$ each time that $j$ would
violate the constraint $b$ assuming the value $s$ (a set of messages
corresponds then to a cluster of configurations).  For locally
tree-like factor graphs (like the ones associated typically to
randomly generated instances) the messages incoming to $j\in
V(a)\setminus i$ can be assumed uncorrelated, after the temporary
removal of a single clause $a$ and of a variable $i$ (the so called
cavity step, see Fig.~\ref{1}).  It becomes then possible to evaluate
probability density functions for the messages, 
called \textit{cavity surveys} (the probability space originates from
the set of all clusters of satisfying assignments sampled with uniform
measure):
\begin{equation}\label{cavitysurvey}
Q_{b, j}(\vec{u}_{b, j})=\eta^0_{b, j}
\delta(\vec{u}_{b, j},\vec{0})+
\sum_{s=1}^q \eta_{b, j}^s\delta(\vec{u}_{b, j}, \vec{e}_s)
\end{equation}
Here, $\eta_{b, j}^s$ are the probabilities that $b$ constrains $j$ not to enter 
the $s$ state
and $\eta_{b, j}^0$
is the probability that no message is sent, $b$ being already satisfied 
by the assignment of other variables. 

The cavity surveys form a closed system of $K M$ functional equations
for which a solution can be found in linear time by iteration \cite{SP, BMZ}:
\begin{equation}\label{convolution}
Q_{a, i}(\vec{u}_{a, i})=\int \mathcal{D}Q\,\,\,
\delta_\mathcal{E}\left[\{\vec{u}_{b, j}\}\right]\,\chi\left[\vec{u}_{a, i}, 
\{\vec{u}_{b, j}\}\right]
\end{equation}
where the function $\chi\left(\{\vec{u}_{b, j}\}\right)$ depends on
the specific CSP and $\mathcal{D}Q=\prod_{j \in V(a)\setminus
  i}\prod_{b\in V(j)\setminus a} Q_{b, j}(\vec{u}_{b, j})$.  The
functional $\delta_\mathcal{E}\left[\{\vec{u}_{b, j}\}\right]$ acts as
a filter, assigning null weight to sets of messages associated to
clusters of excited configurations.  At the fixed point, from the
knowledge of the surveys one may compute the fractions $W_{j}^s$
($W_{j}^0$) of clusters of solutions in which a variable $j$ is frozen
in the direction $s$ (or is unfrozen). Such microscopic information
can be successfully used to find optimal assignments by decimation
\cite{BMZ}.

We shall now present a generalization of the SP algorithm (SP-ext),
allowing the retrieval of a solution close to any desired
configuration $\vec{\xi}$.  Hereafter we shall refer for simplicity to
the $K$-SAT problem ($s=\pm 1$ only), but the method could be easily
extended to a generic CSP.  On general grounds, a way to analyze
specific regions of the configuration space would be to solve the
cavity equations in presence of an additional field conjugated to some
geometrical constraint (\textit{e.g.} fixed magnetization). This
strategy would be however algorithmically inefficient in that it would
change the nature of the components of the messages (from integers to
reals), slowing down significantly the iterative solution of the SP
equations.

Here we consider instead an arbitrary but quite natural extension of
the SP equations in which
external messages $\vec{u}_i=\vec{e}_{\mbox{{\tiny -}}\xi_i}$
(represented in Fig.~\ref{1} as triangles) in an arbitrary direction
$\vec{\xi}\in\{-1, 1\}^N$ are introduced for each variable.  New
associated surveys
\mbox{$Q_{i}^\pi(\vec{u}_i)=(1-\pi)\delta(\vec{u}_i, \vec{0})+
  \pi\delta(\vec{u}_i, \vec{e}_{\mbox{{\tiny -}}\xi_i})$} are given
\textit{a priori} and never updated, and affect dynamically the
relative weight of the different clusters, entering into the measure
$\mathcal{D}Q$ in the convolution integrals (\ref{convolution}).  The
parameter $\pi$ can be interpreted as strength of the perturbation.
Convergence can be reached only if the zero energy constraint is
respected.  While an intensity $\pi\simeq 1$ would produce a complete
polarization of the messages if $\vec{\xi}$ was a solution, in the
general case the use of a smaller forcing intensity allows the system
to react to the contradictory driving and to converge to a set of
surveys sufficiently biased in the desired direction, allowing for an
efficient selective exploration of specific parts of the solution
space.

In order to use SP-ext for probing the local geometry of the zero
energy phase one proceeds as follows. First, a random solution
$\vec{\sigma}$ is found by decimation (SP-ext, differently from the
standard SP \cite{BMZ}, is typically able to retrieve complete
solutions).  Next, new satisfying assignments $\vec{\sigma}_\delta$
are generated, forcing now the system along a direction obtained
flipping $N\delta$ spins of the original solution $\sigma$.  In order
to have a highly homogeneous distribution of the clusters, we have
chosen for our experiments an ensemble of random K-SAT in which
variables have fixed degree and are balanced (i.e. have an equal
number of directed and negated occurrences in the clauses) For this
specific ensemble and for $K=5$, one has $\alpha_d= 14.8$ and
$\alpha_c= 19.53$.  Between $\alpha_d$ and $\alpha_G=16.77$ the phase
is expected to have multiple levels of clustering (a full RSB phase
\cite{MMZ, MoPR}) whereas between $\alpha_G$ and $\alpha_c$ the 1-RSB
phase is stable.  We have estimated $\alpha_G$ with a new message
passing algorithm implementing the cavity equations at the level of
two steps of RSB \cite{baldassi}.

\begin{figure}
\begin{centering}
\includegraphics[scale=0.4835]{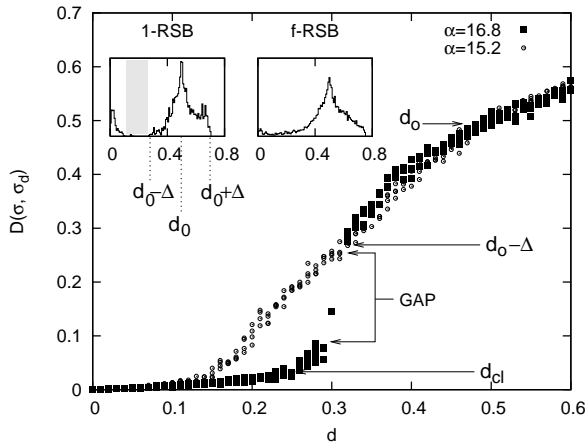}
\caption{Distribution of distances}\label{2}
\end{centering}
\end{figure}
In the experiments we have taken instances of size $N=10^4$ with an
intensity $\pi=.35$ of the forcing (close to the highest value of
$\pi$ for which the SP-ext equations always converged for this sample).  
The Hamming
distance between $\sigma$ and $\sigma_\delta$ is plotted 
against $\delta$ in the first \textit{stability diagram} (black
data points) shown in Fig.~\ref{2}.  For $\delta < \delta_c\simeq
0.3$, $d_H(\vec\sigma,\vec\zeta_\delta)$ linearly increases with a
very small slope until a value $d_{\mbox{\tiny{cl}}}$.  Conversely,
for $\delta > \delta_c$, it jumps to a value $d_0-\Delta$, and a
symmetric distribution of distances around $d_0$ is obtained.  Under
the hypothesis of homogeneous distribution of clusters, the fixed
point average site magnetization \mbox{$\langle W^+ -W^-\rangle$}
provides an analytic estimation of the typical overlap
$q_0=\frac{1-d_0}{2}$ between two different clusters in agreement with
the experiments. On the other hand, $d_{\mbox{\tiny{cl}}}$ is of the
order of the average fraction of unfrozen variables \mbox{$\langle W^0
  \rangle$}. The gap between clusters is the main prediction of the
1-RSB cavity theory which finds in these experiments a nice confirm.

A completely different behavior is observed when repeating the
experiment in the expected full RSB phase.  The histogram of the
reciprocal distances among all the generated solutions (inset of
Fig.~\ref{2}, gray bars) is now gapless, unlike the previous sample
(black bars), and the related stability plot in the main figure (white
data-points) deviates significantly from the 1-RSB case.  Various
shapes of the stability plot, differing for their degree of convexity,
can be obtained starting from different solutions of the same sample.
This hints to the existence of a mixed phase, in which higher order
hierarchical clustering is present and in which many local cluster
distribution topologies can coexist.

\begin{figure}
\begin{centering}
\includegraphics[scale=0.4835]{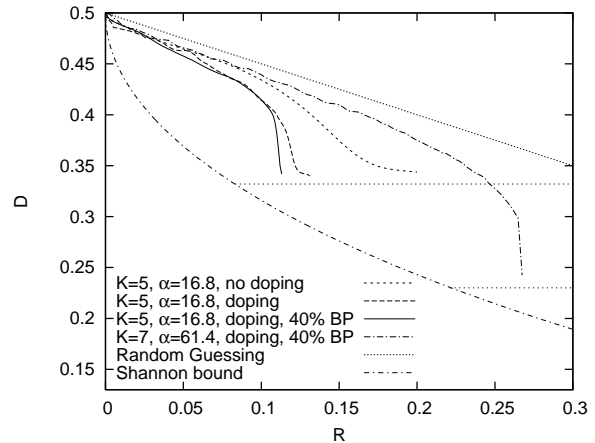}
\caption{Rate-Distortion profile for the compression of a random unbiased 
source}\label{3}
\end{centering}
\end{figure}
The ability of SP-ext to select clusters is a new computational
feature which may play a role in different systems in which clustering
of input data is important.  Here we consider a first basic
applications by implementing a lossy data compressor \cite{mackay}
which exploits the 1-RSB clustered structure for data quantization
purposes.

Let us suppose to have an input $N$-bit binary string $\vec{\xi}$
generated from an unbiased and uncorrelated random source.  Given an
appropriate $K$-SAT instance with $N$ variables, a solution
$\vec{\sigma}_{\xi}$ as close as possible to $\vec{\xi}$ can be
generated with SP-ext.  One can expect to find a solution at a
distance close to \mbox{$d_0 - \Delta$}, if the cluster distribution
is homogeneous (balanced and fixed even connectivity instances are
then chosen). Furthermore, $\alpha$ is taken slightly larger than
$\alpha_G$, in order to maximize the number of addressable clusters,
still preserving a sharp separation among them.
At this point, a compressed string $\vec{\sigma}_{\xi}^R$ is built by
retaining just the spins of the first $NR$ variables of
$\vec{\sigma}_{\xi}$.  In the decompression stage, SP-ext is run over
the same graph, applying a very intense forcing ($\pi=0.99$) parallel
to $\vec{\sigma}_{\xi}^R$.  If $R>R_c$, SP-ext becomes able to exactly select the
single cluster to which $\vec{\sigma}_{\xi}$ belongs.  The cluster
addressing is actually so sharp, that no decimation is needed and all
the remaining $N(1-R)$ unforced variables can simultaneously be fixed
to their preferred orientation without creating contradictions.  A
comparison with the theoretical Shannon Bound \cite{mackay} is done in
Fig.~\ref{3} (dotted line), where the cluster selection transition is
clearly visible. The accumulated distortion with respect to
$\vec{\xi}$ is of the order of $d_{\mbox{\tiny{cl}}}+d_0-\Delta$ (the
horizontal lines in Fig.~\ref{3} refer to the original distance
between $\vec{\xi}$ and $\vec{\sigma}_\xi$, and, obviously, no better
distortion can be achieved in this scheme).  The line relative 
to the performance of a trivial decoder in which the
missing bits are randomly guessed is also plotted.

The Shannon Bound can be approached by the use of \textit{iterative
  doping} \cite{verdu} technique for chosing the bits to store.  After
the determination of $\vec{\sigma}_\xi$, SP-ext is run again
\textit{without} applying any forcing and a ranking of the most
balanced variables is performed. One looks for the variable $i$ which
minimizes $|W_i^+-W_i^-|+W_i^0$ (frozen in opposite directions in a
similar number of clusters and rarely unconstrained).  The state
assumed by $i$ in the solution $\vec{\sigma}_{\xi}$ will be taken as
the first bit of the compressed string $\vec{\sigma}_{\xi}^{R}$ and
used to fix $i$.  New doping steps are done, until when the desired
compression rate has been reached.  An \textit{identical} doping stage
is then performed in decompression.  The iterative ranking allows
indeed to find out which spins have to be fixed accordingly to the
ordered bit sequence $\vec{\sigma}_{\xi}^{R}$ and the left variables
can be fixed as in the previous decompression method.  Fixing a
balanced variable ``switches off'' a larger number of clusters, and,
hence, the critical rate is reduced (dash dotted curve in
Fig.~\ref{3}), thanks to a less redundant coding of the information
needed for the cluster selection.  A further improvement can be
obtained, by using in the doping stage a modified iteration, that
interpolates between SP and the well known Belief Propagation
\cite{BP} equations (solid curve in Fig.~\ref{3}).

 \begin{figure}
\begin{centering}
\includegraphics[scale=0.4835]{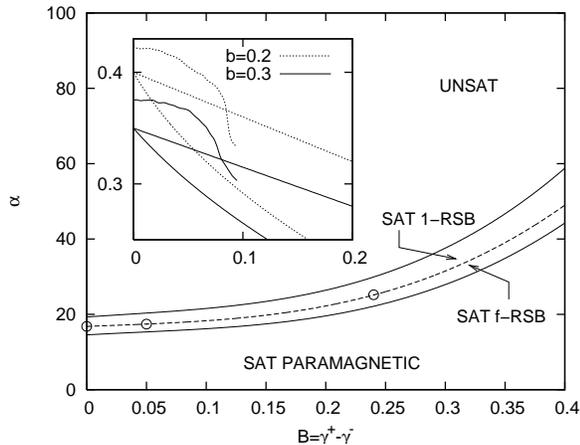}
\caption{Phase diagram for fixed degree biased 5-SAT and compression performance 
for 
two biased sources.}\label{4}
\end{centering}
\end{figure}
SP-ext can also be applied when the occurrence probability $P_\pm$ of
the possible input symbols are different. The issue is of practical
relevance since correlated sources can often be shown to be equivalent
to memoryless biased sources \cite{BWT}.  Let suppose that the bias of
the source is $b= P_+-P_- > 0$.  It is possible to engineer graphs
with a cluster distribution concentrated around the ferromagnetic
direction, by making for every variable $i$ the fraction $\gamma_+$ of
couplings $J_{a,i}=+1$ larger than the fraction $\gamma_-$ of
$J_{a,i}=-1$.  As shown indeed in Fig.~\ref{4}, a narrow SAT RSB
stripe is still present for a balancing $B=\gamma_+-\gamma_- < 0.435$.
When the balancing is too large, there is on the other hand a direct
transition between an unfrozen SAT phase and the UNSAT region (the
line where optimized heuristics start to fail in determining a
solution approximately continues the 1-RSB SAT/UNSAT transition line).
The rate-distortion profile of the compression of a random
uncorrelated source with $b=0.2$ and $b=0.3$ is shown in the inset of
the same figure.  The best found graphs for all the analyzed values of
$b$ are always located in proximity of the Gardner line (empty circles
in Fig.~\ref{4}, better critical rates but worse distortions are found
for higher connectivities).  Cluster selection is still possible and
ensures again a non-random correlation between the input and the
output string. It is expected nevertheless that much better
performances can be achieved by a careful optimization of the graph
ensemble as it was shown for iterative decoding with Belief
Propagation \cite{graph_optimization}.

We conclude by noticing that work is in progress in order to obtain a
fully local version of the decimation procedure leading to the
different solutions, by mean of a local reinforcement technique: this
fact might be important for the parallelization of the SP algorithm
and for modeling distributed computation in complex networks.

Acknowledgements. This work has been supported by the EVERGROW EU project.


\begin{thebibliography}{99}

\bibitem{Garey_Johnson}{M.R. Garey, D.S. Johnson, 
\textit{Computers and Intractability}, Freeman, New York (1979)}

\bibitem{Nature} R. Monasson, R. Zecchina, S. Kirkpatrick, B. Selman,
and L. Troyanski, Nature 400, 133  (1999)

\bibitem{TCS} Special Issue on {\it NP-hardness and Phase transitionls}, 
edited by O. Dubois, R. Monasson, B. Selman and R. Zecchina, 
Theor. Comp. Sci. 265, Issue: 1-2 (2001)

\bibitem{SP} M. M\'ezard, G. Parisi, R. Zecchina, Science 297, 812
(2002); M. M\'ezard, R. Zecchina, Phys. Rev. E 66, 056126
(2002)

\bibitem{BMZ} A. Braunstein, M. M\'ezard, R. Zecchina, Survey
propagation: an algorithm for satisfiability, preprint, to
appear in {\it Random Structures and Algorithms}, cs.CC/0212002

\bibitem{LDPC} T.~Richardson, R.~Urbanke,
An introduction to the analysis of iterative coding systems,
in {\it Codes, Systems, and Graphical Models}, edited by
B.~Marcus and J.~Rosenthal, Springer, New York, 2001

\bibitem{mackay}{D.J.C. MacKay, Information Theory, Inference, and Learning 
Algorithms, Cambridge University Press, Cambridge, MA (2003)}

\bibitem{coloring} A. Braunstein, R. Mulet, A. Pagnani, M. Weigt,
R. Zecchina, Phys. Rev  E 68, 036702 (2003)

\bibitem{MMZ} S.Mertens, M. M\'ezard, R. Zecchina, Threshold values of
Random K-SAT from the cavity method, preprint, cs.CC/0309020 (2003)

\bibitem{MoPR} A. Montanari, G. Parisi, and F. Ricci-Tersenghi, J. Phys. A 37, 
2073 (2004)

\bibitem{kirkpatrick:selman:94} S. Kirkpatrick, B. Selman, Science 264, 1297 
(1994)

\bibitem{baldassi} C. Baldassi, R. Zecchina, to appear

\bibitem{verdu} G.Caire, S. Shamai, S.Verd\`u, A new data compression algorithm 
for sources with memory based on error correcting codes, in proc. IEEE Information 
Theory Workshop, 
pp. 291--295, Paris, (2003)

\bibitem{BP} J.S. Yedidia, W.T. Freeman and Y. Weiss,
Generalized Belief Propagation, in {\it Advances in Neural Information
Processing Systems 13} , 689--695, MIT Press (2001)

\bibitem{BWT} M. Burrows, Tech. Rep. SRC 124 (1994)

\bibitem{graph_optimization}{D.J.C. MacKay, R.M. Neal, IEEE
Electronics Letters, 32, 18, 1645-1655 (1996);  S.Y. Chung, G.D. Forney,
T.J. Richardson, R. Urbanke, IEEE Communications Letters, vol. 5,
no. 2, 58 - 60 (2001)}

\end{thebibliography}
\end{document}